\documentclass{mem0}
\usepackage{natbib}\usepackage{txfonts}\usepackage{balance}
\usepackage{graphicx}
\usepackage[a4paper]{hyperref}
\idline{75}{282}
\begin{document}
\def\mum{$\mu$m~}
\def\lH{$l/H$}
\def\Vt{$V_t$}
\def\CDC{$^{12}$C/$^{13}$C~}
\def\HHO{H$_2$O~}
\def\Tef{T$_{\rm eff}$}
\newcommand{\ai}{{\em ab initio}}

\def\teff{$T\rm_{eff }$}
\def\kms{$\mathrm {km s}^{-1}$}

\title{The modelling of the spectra and atmospheres of evolved stars
}

   \subtitle{}

\author{
Yakiv V.Pavlenko\inst{1}
          }

  \offprints{Ya. V.Pavlenko}

\institute{
Main Astronomical Observatory of the National Academy of Sciences,
27 Zabolotnoho, Kyiv-127, 03680 Ukraine
\email{yp@mao.kiev.ua}
}

\authorrunning{Pavlenko}

\titlerunning{Model atmospheres and spectra}

\abstract{

The method and results of the computation of the model atmospheres
and spectral energy distributions of chemically peculiar stars, are discussed.
The models are computed with a special consideration of the particular
problems encountered when computing model atmospheres for 
M and C-giants, and of hydrogen deficient stars. 
We present some computed model atmospheres for 
Sakurai's object, giants of globular clusters, and C-giants.
}
\maketitle{}

\section{Introduction}

The computations of model atmospheres of and theoretical spectra
of stars is an essential part of many modern astrophysical
projects. An extended grid of model atmospheres and spectra were
computed recently using the most complete sets of opacities (see
Kurucz (1993, 1999), Hauschildt et al. (1999). In most of the computations
the solar abundances (Anders \& Grevesse 1989) 
or solar abundances scaled by the metallicity factor [Fe/H]
are used. However,
the metal abundances in the atmospheres of evolved stars can 
significantly differ from the solar abundance ratios. The reason for this is 
because convection and other mixing processes dredge up
the products of the nucleosynthesis from the stellar interior.
%% Molecular opacities are of importance in the
%%atmospheres of C- and M-giants. In the case of coolest stellar atmospheres
%%(\Tef $<$ 3000 K)
%%we should account absorption of polyatomic species and dust.

The atmospheres of the most evolved stars (R CrB, Sakurai's object, etc) present
another interesting problem. They are helium and carbon rich, therefore the
opacity due to H$^-$ absorption is not as important.
The temperature structure of the model atmospheres of these stars
is the different from the case of the ``normall''  abundances 
due to the changes of opacity.

\section{Model atmospheres and synthetic spectra}

We computed plane-parallel model atmospheres
of the evolved stars in LTE, with no energy
divergence by SAM12 program (Pavlenko 2003). The
program is a modification of ATLAS12 (Kurucz 1999).

Chemical equilibrium is computed for the molecular species, 
by assuming LTE. The nomenclatures of molecules accounted for are
different in atmospheres of hydrogen rich and hydrogen poor,
carbon-rich and carbon poor stars. We account mainly for the molecules
which are the most abundant or most important sources of opacities.

SAM12 uses the standard set of continuum opacities from ATLAS12.
The adopted
opacity sources account for changes in the opacity as
a function of temperature and element abundance.
We add some opacities sources which are of importance in the atmospheres of
carbon-rich, hydrogen-deficient stars:

-- The opacities due to
C~I, N~I, O~I bound-free absorption over a wide (0.1--8~$\mu$m)
wavelength region were
computed using the OPACITY PROJECT (Seaton et al.1992) cross sections database
(Pavlenko \& Zhukovska 2004).

-- The opacity of C$^-$ (Myerscough \& McDowell 1996) also is
taken into account
(see Pavlenko 1999 for more details).

The opacity sampling approach (Sneden et al. 1976) is used to account for
atomic and molecular line absorption. The atomic line data are taken
from the VALD database (Kupka et al. 1999). Lists of diatomic
molecular lines of $^{12}$CN, $^{13}$CN, $^{12}$C$_2$, $^{13}$C$_2$,
$^{12}$C$^{13}$C, $^{12}$CO, $^{13}$CO, SiH, and MgH are taken from
Kurucz (1993). We account for
TiO and VO opacities in the atmospheres of oxygen-rich stars in the
framework of JOLA (just-overlapping-line) approximation. In the atmospheres of
the carbon-rich stars the absorption by CS lines (Chandra et al. 1994)
was accounted for.

Bands of polyatomic molecules appear in the spectra of the coolest
evolved stars:
%%. We account for them using a different method than
%%for the computation of model atmospheres of M- and C-giants:

a) Water vapour band form strong features in the infrared spectra
of cool stars with oxygen-rich atmospheres. We computed \HHO
opacities using AMES line lists (Partrige \& Schwenke 1998).

b) In the cool atmospheres of C-stars HCN/HNC bands dominate in the infra-red
region of the spectrum. We computed the HCN/HNC opacity using
detailed line line list (Harris et al. 2002, 2003, 2006).

The shape of each molecular or atomic line was determined using
the Voigt function $H(a,v)$. Damping constants were taken from
line databases or computed using Unsold's (1955) approach.

The following molecular electronic bands
were accounted for:
CaO (C$^1\Sigma$ - X$^1\Sigma$),  CS(A$^1\Sigma$ - X$^1\Sigma$),
SO (A$^3\Pi$ -  X$^3\Sigma$), SiO (E$^1\Sigma$-X$^1\Sigma$),
SiO(A$^1\Pi$ -  X$^1\Sigma^+$, NO (C$_2 \Pi_r$- X$_2\Pi_r$),
NO(B$_2\Pi_r$ -X$_2\Pi_r$),
NO(A$^2\Sigma^+$ - X$_2\Pi_r$),
MgO(B$^1\Sigma^+$ -  X$^1\Sigma^+$),
AlO(C$^2\Pi$ -X$^2\Sigma$),
AlO(B$^2\Sigma^+$ - X$^2\Sigma^+$),
in the framework
of the JOLA approach.

Convection plays an important role in
these atmospheres.
We use the mixing length theory of 1D convection modified by
Kurucz(1999) in ATLAS12.

\begin{figure*}
\includegraphics[width=130mm]{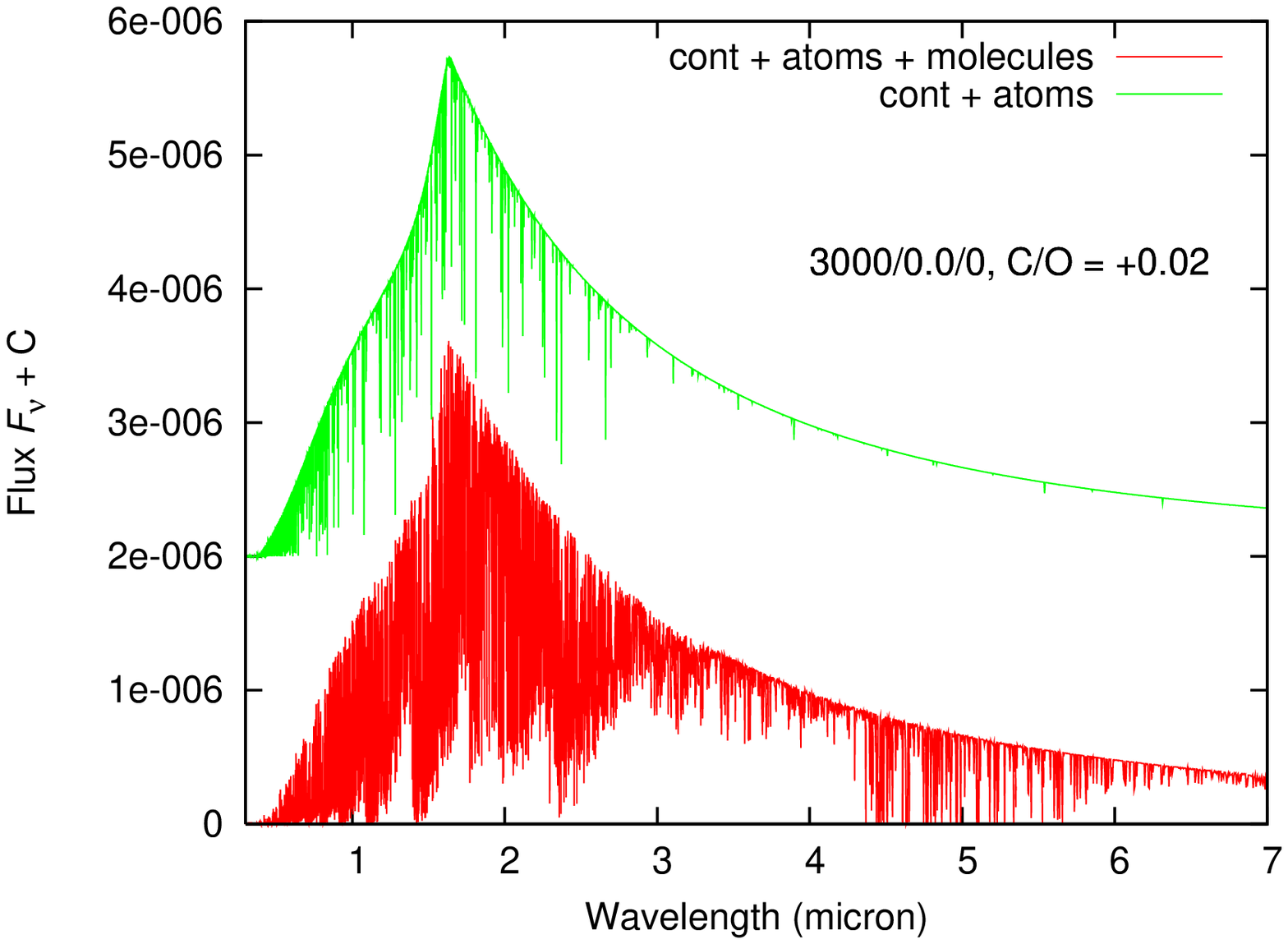}
\includegraphics[width=130mm]{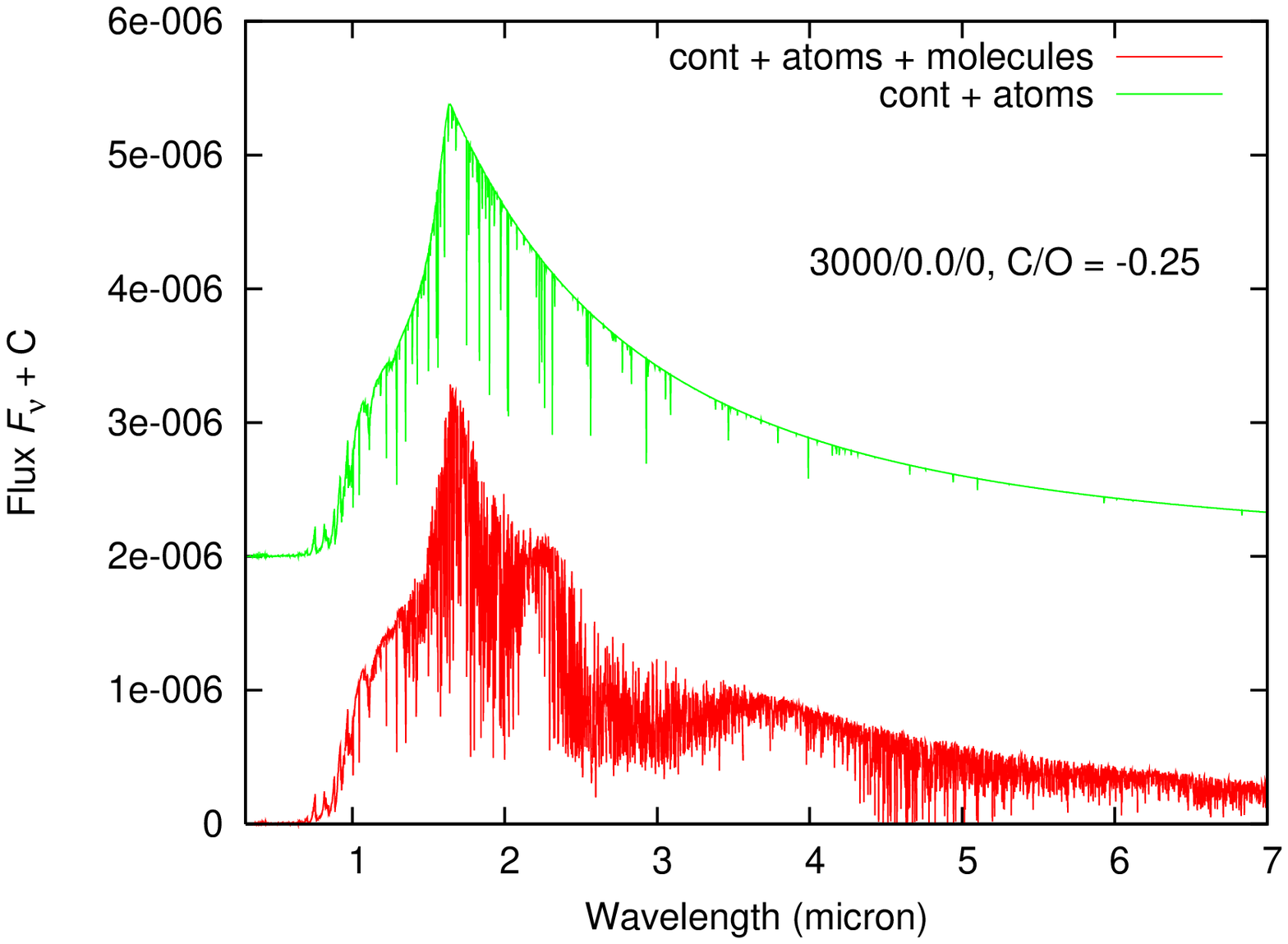}
\caption{\label{_SEDS}Spectral energy distributions computed for
two giants with \Tef/log g/[Fe/H] = 3000~K/0.0/0 and different C/0
= -0.25 (top) and (+0.02) (bottom). }
\end{figure*}

In the Fig. \ref{_SEDS} we show the computed SEDS for two model atmospheres
of the same effective temperature, surface gravity and metallicity, 
i.e 3000/0.0/0,
but with a different C to O ratios. 
The differences in the opacity nomenclatures are clearly seen.

\begin{figure*}
\includegraphics[width=65mm]{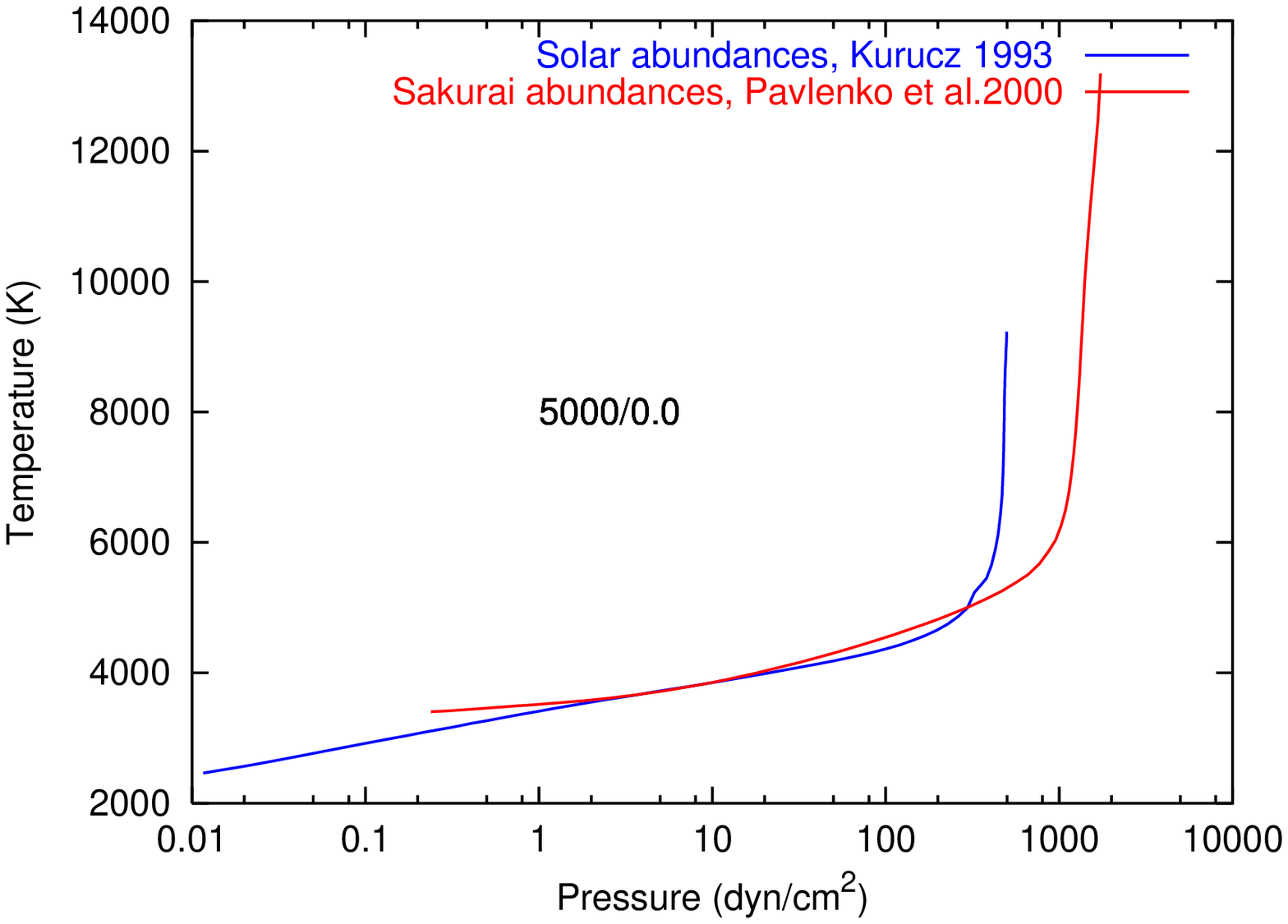}
\includegraphics[width=65mm]{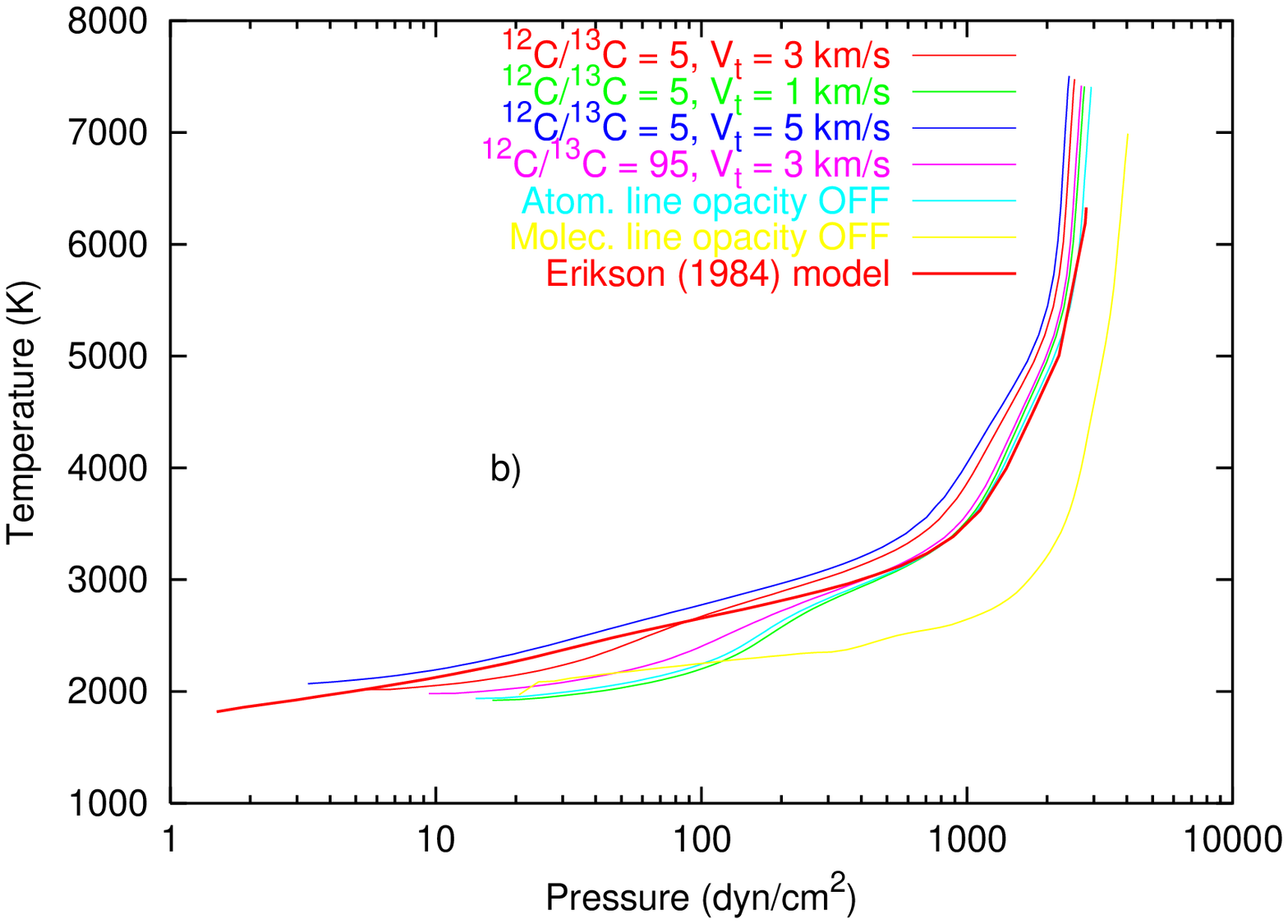}
\caption{\label{TT}Left: temperature structure of the model
atmosphere of giant 5000/0.0 computed with a normal abundance and
with a hydrogen deficient abundance, i.e. [H]=0, -2.4 (see section \ref{Sak}). Right: the
response of the atmosphere temperature of a C-giant 3000/0.0/0 to the
variation of some input parameters.}
\end{figure*}

The temperature structure of model atmospheres of the evolved giants show very
strong dependence on the adopted abundances and other parameters
(Fig. \ref{TT}).

Synthetic spectra are calculated with the WITA6 program
(Pavlenko 2000),
using the same approximations and opacities as SAM12. In the computations we
use TiO line list (Plez 1998) instead of JOLA, and the CO (Goorvitch 1994)
line list instead of the Kurucz(1993) line list. Naturally, much smaller
wavelength steps in the synthetic
spectra are adopted ($\Delta \lambda$ = 0.02 and 0.1 in the blue 
and IR spectral regions, respectively.

\section{Fits to observed spectra}
\subsection{\label{_vfc}Veiling-free case}

To determine the best fit parameters, we compare the observed
fluxes $F_{\nu}$ with the computed fluxes
following the scheme of Jones et al (2002) and Pavlenko \& Jones (2003). We
let
$$F_{\nu}^x = \int F^y_{\nu} \times G (x-y)*dy, $$
where $r^y_{\nu}$ and $G(x-y)$ are respectively the
fluxes computed by WITA6 and the broadening profile. We adopt
a gaussian + rotation and/or extension profile for the latter.
We then find the minima of the 3D
function
 $$S(f_{\rm s}, f_{\rm h}, f_{\rm g}) =
   \sum \left ( F_{obs} - F^x \right )^2  , $$
where $f_{\rm s}$, $f_{\rm g}$, $f_{\rm g}$ are the
wavelength shift, the normalisation factor, and the profile
 broadening parameter, respectively.
The parameters $f_{\rm s}, f_{\rm h}$ and $f_{\rm g}$
are determined by the minimisation procedure for every computed
spectrum. Then, 
From the grid of the better solutions for the given
abundances and/or other parameters (microrurbulent velocity,
effective temperature, isotopic ratios, etc), we choose the
best-fitting solution.

\subsection{\label{_vc}A case with the dust veiling}

For the case of the most evolved stars, some of the flux originates in the
optically thin envelope around the star. Third provides a
source of veiling the observed spectra: $F_{total} = F_{atmos}+F_{envelope}$. 
In this case we should minimise $F_{total}-F_{obs}$, whilst 
considering $F_{envelope}$
as an additional parameter (see section \ref {Sak} and Pavlenko et al. 2004
for more details).

\section{Results}

\subsection{C-giants}

The {\it ab initio} HCN linelist (Harris et al. 2002, 2003) was
used to compute new model atmospheres and synthetic spectra for
the carbon stars (Pavlenko 2003). Recently the accuracy of the first {\it
ab initio} line list was improved by by using an accurate
database of 5200 HCN and HNC rotation-vibration energy levels,
determined from existing laboratory data.

Model atmospheres and synthetic
spectra computed with this new linelist provide better fit to 
the spectrum of WZ~Cas in which the
absorption feature at 3.56 \mum\  is reproduced to a higher degree
of accuracy than has previously been possible (see Harris et al,
2006 for more details).

Using the new HCN line list, we computed a new grid of carbon rich
model atmospheres for a wide range of [Fe/H], C/O, \Tef, log g.
They can be used for the express analysis of the new data obtained
for C-giants in our Galaxy (Yakovina et al. 2003)
and/or other galaxies (Dominguez et al.
2005). 

\subsection{\CDC in giants of globular clusters}

\begin{figure*}
\begin{center}
\includegraphics[width=65mm,angle=00]{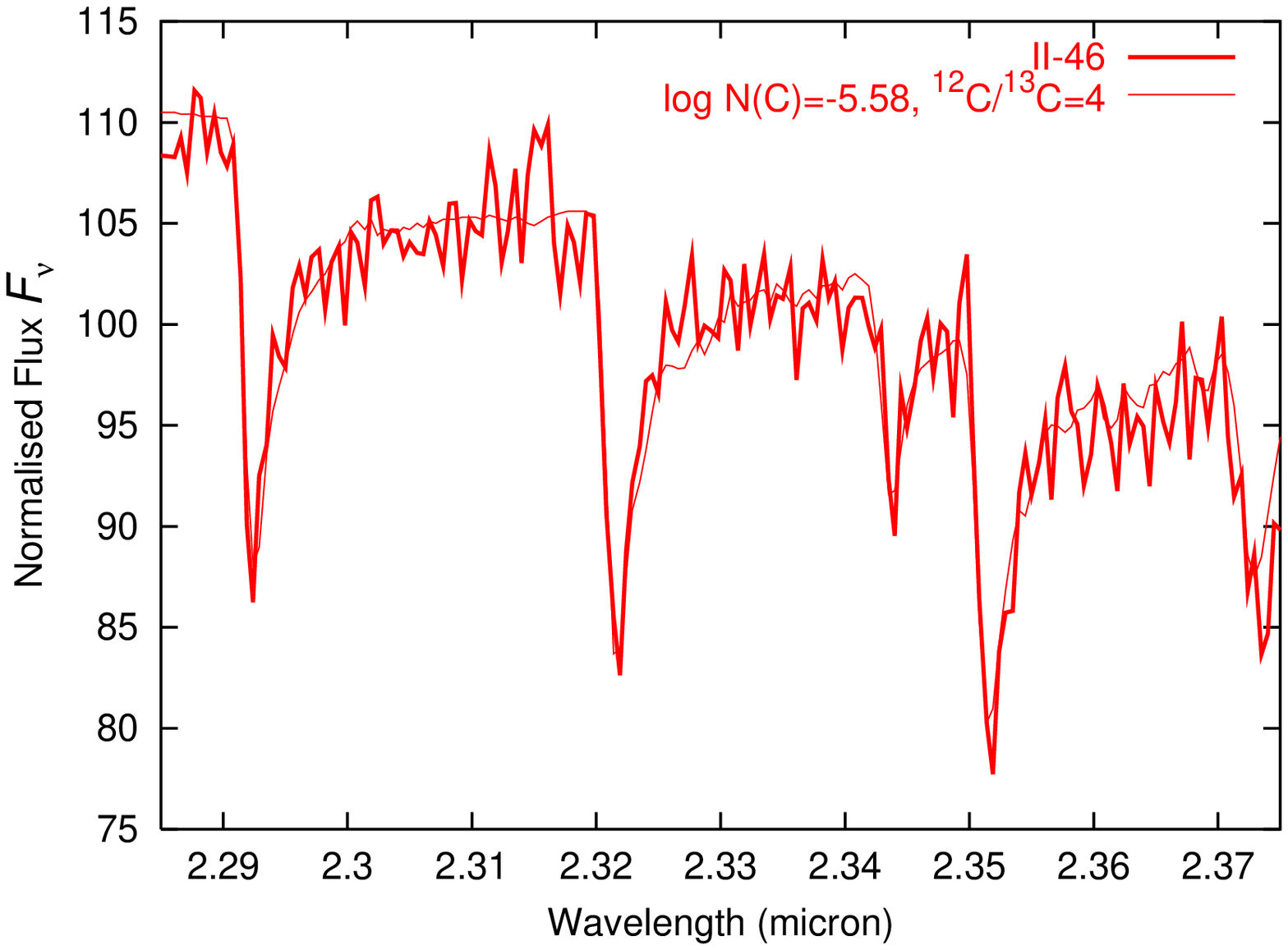}
\includegraphics[width=65mm,angle=00]{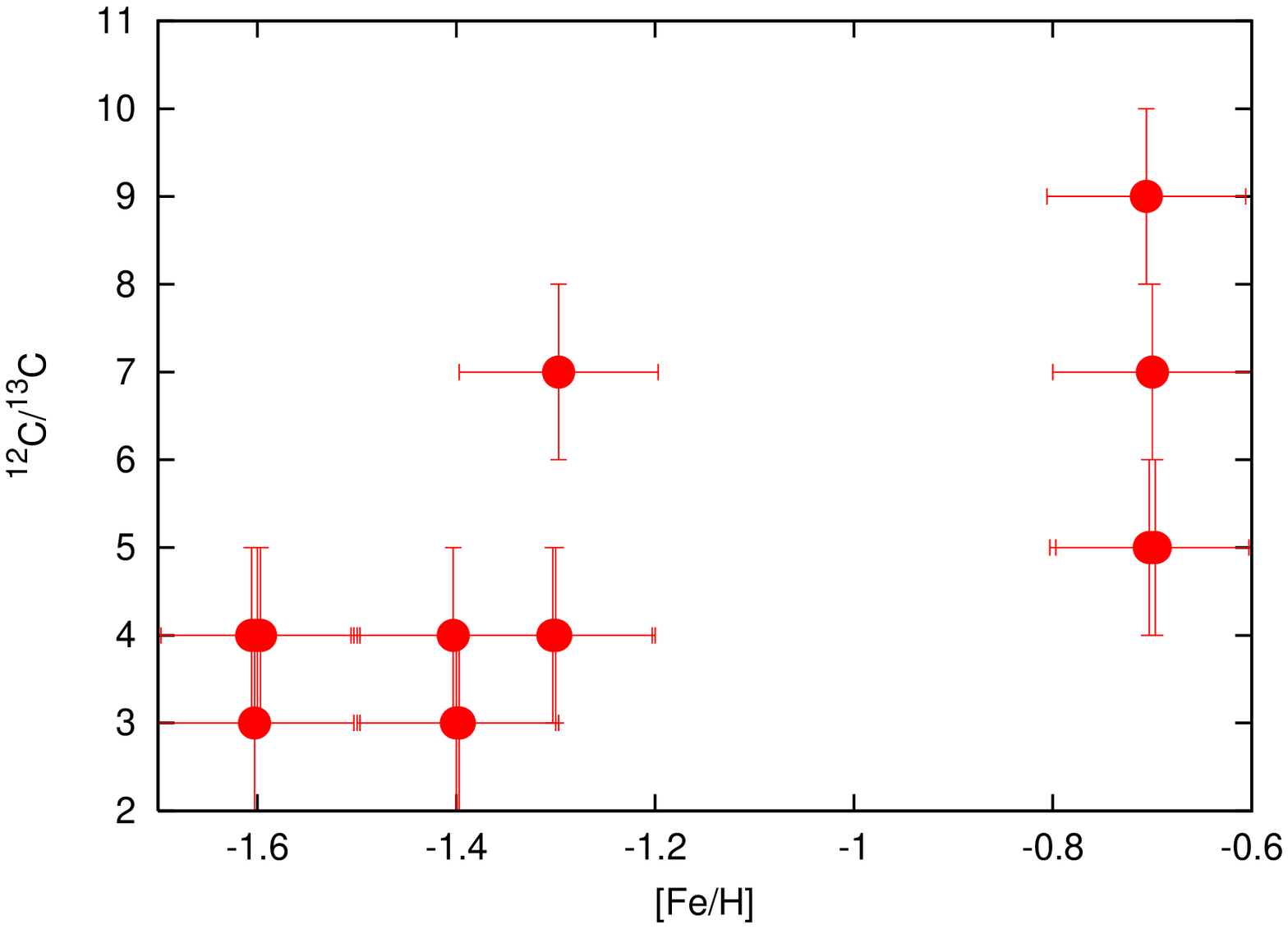}
\end{center}
\caption[]{\label{_M} Left: fits to observed spectrum of
II-46 giant in the globular cluster M3. Right: the found \CDC for
giants of globular clusters of different metallicities (Pavlenko et al. 2003b).}
\end{figure*}

%The behaviour were examined by
%comparing observed and synthetic spectra of
%stars in globular clusters of different metallicities (
Pavlenko, Jones
\& Longmore (2003b) investigated 
changes in the $^{12}$C/$^{13}$C isotopic ratio and the carbon abundances
in atmospheres of red giants of  in the galactic globular
clusters M71, M5, M3 and M13 covering the metallicity range
from --0.7 to --1.6 and \Tef = 3500--4900 K. Observational data were 
obtained with the UKIRT telescope in 1993.
Model atmospheres and synthetic spectra of the $\Delta\nu$=2 CO bands 
around 2.3 $\mu$m 
were computed for a fixed
\Tef and log g, but with different \Vt, C/O, \CDC. We then apply the minimisation
procedure (see sect. \ref{_vfc}) to obtain the best solution. An example of the best fit
 to an observed spectrum of II-46 giant of globular cluster M3 is shown in Fig.
 \ref{_M}.
We find:\\
-- lower \CDC for more luminous hotter objects. \\
-- relatively low carbon abundances which
are not affected by the value of oxygen abundance. \\
-- For
most giants the determined $^{12}$C/$^{13}$C ratios are consistent with the
equilibrium value for the CN cycle. \\
-- a larger dispersion of
\CDC in giants of M71 of metallicity [M/H]
= -0.7 in comparison with other
giants of M3, M5, M13 which are more metal deficient (Fig. \ref{_M}).

%%%% Figure

Our results suggest a complete mixing on the
ascent  of the stars up the red giant branch. This is in 
contrast to the substantially higher values
suggested across this range of parameters by the current
generation of theoretical models of stellar evolution
(see Pavlenko et al. 2003b for more details).

\subsection{\label{Sak}R CrB and V4334 Sgr}

It has been firmly established that the two most abundant elements in
the atmospheres of the evolved R CrB-like stars are helium and
carbon (see Asplund et al 1997).
A hydrogen deficient model
atmosphere for R CrB was computed by Pavlenko (1999), see
www.mao.kiev.ua\-/staff\-/yp/\-Results\-/Mod.rcrb1999.tar.gz.

Sakurai's object (V4334 Sgr)
 provides another extreme case of stellar evolution.
The ``novalike object in Sagittarius'', was discovered by Y.~Sakurai
on February 20, 1996 (Nakano et al. 1996). Soon after discovery, it was
found to be a final He flash object, a rare type of star on 
the evolutionary track
that leads from the central star of a planetary nebulae back to the red giant
region. These objects are often referred to as  ``born-again giants''. 
Its progenitor was a
faint blue star ($\sim$ 21$\rm ^m$) in the centre of a low surface brightness
planetary nebula (Duerbeck \& Benetti 1996). 
%%Early spectroscopic studies of
%%the object found an increasing hydrogen deficit in the atmosphere, and a
%%C/O ratio $>$ 1 (Asplund et al. 1997, Kipper \& Klochkova 1997).

\begin{figure*}
\begin{center}
\includegraphics[width=65mm,angle=00]{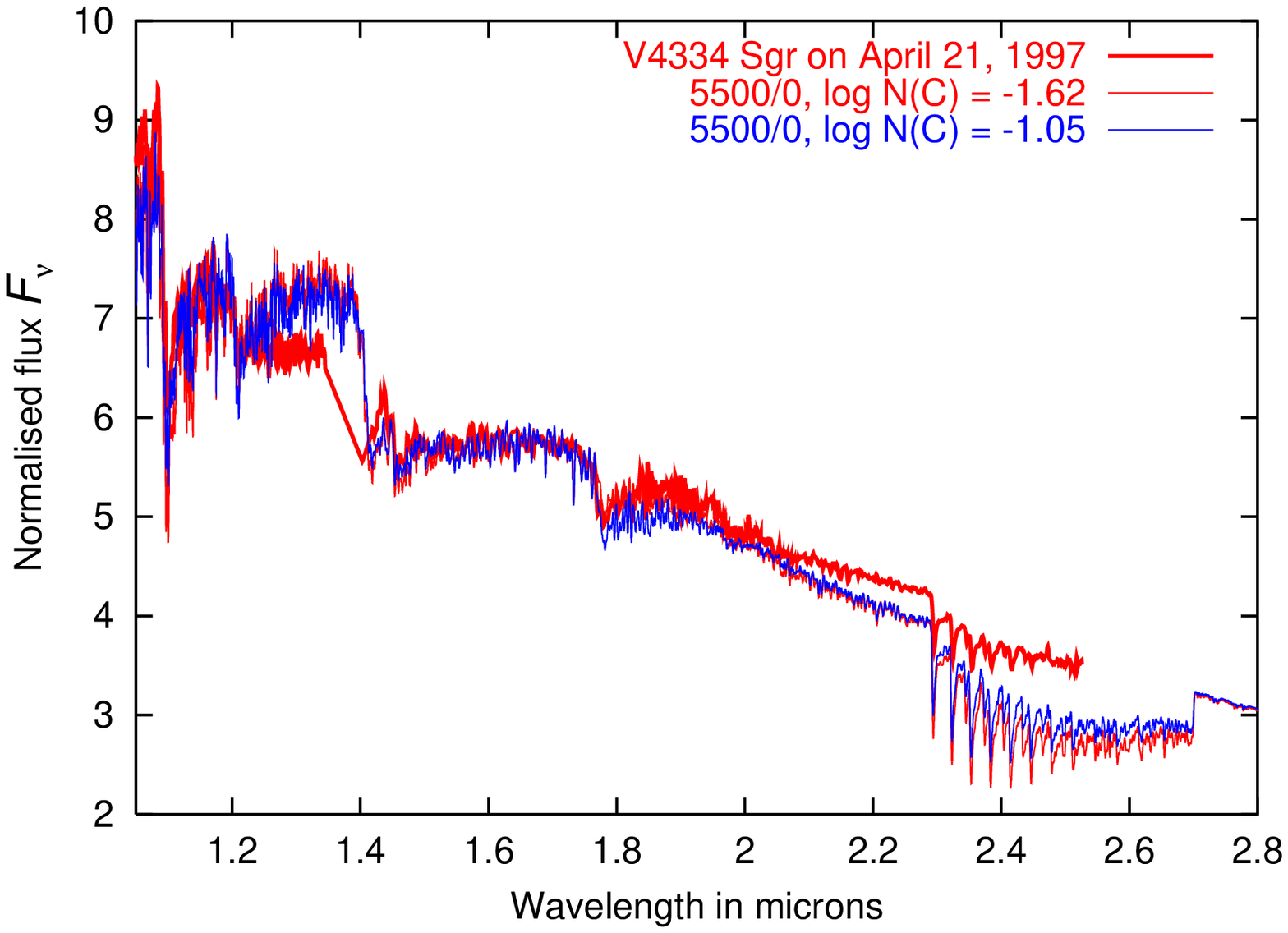}
\includegraphics[width=65mm,angle=00]{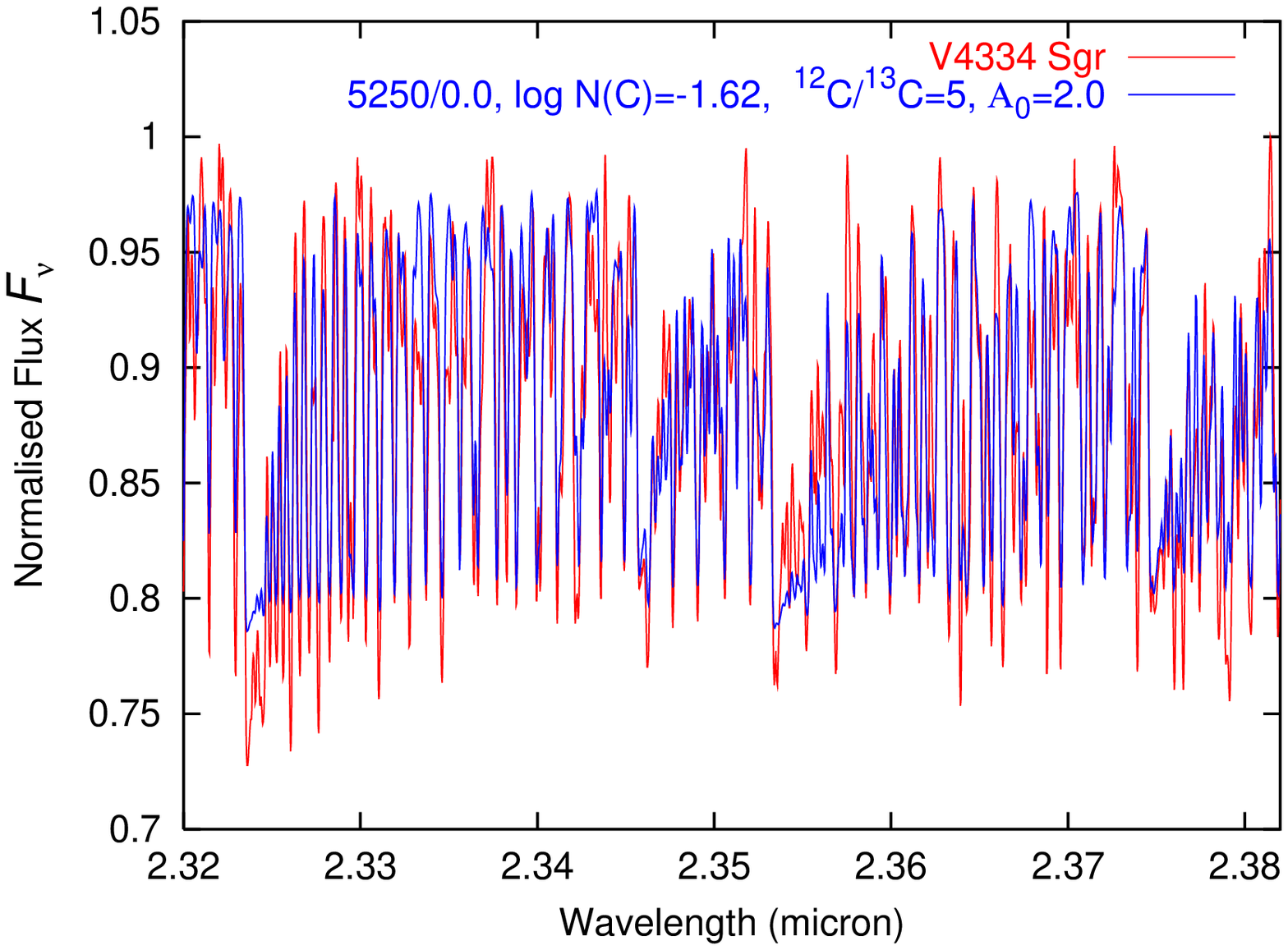}
\end{center}
\caption[]{\label{l6} Left: fits to observed spectrum of of
Sakurai's Object on 1997 July 13. Right: the fit to $^{12}$CO and $^{13}$CO
bands. Synthetic spectra were computed for a microturbulent
velocity of 6~km/s.}
\end{figure*}

Model atmosphere and spectra of Sakurai's object on later stages of evolution
 were investigated
by Pavlenko et al. (2000), Pavlenko
\& Duerbeck (2002). We show the effective temperature of the
pseudo-photosphere of Sakurais object drops monotonically from 5500
in July, 1997 down to 5250 K in August, 1998.
In March 1997 the first evidence of dust formation was seen ( 
Kimeswenger et al. 1997, Kamath \& Ashok 1999, Kerber et al. 2000). 
Pavlenko \& Geballe (2002) {\it spectroscopically~}
showed the presence of the dust veiling of the infrared
part of the Sakurai's spectrum (see Fig. \ref{l6}). Later,
Pavlenko et al. (2004) using the procedure decsribed in
section \ref{_vc}  
determined \CDC = 4 $\pm$ 1 in atmosphere
of Sakyrai's object from the fit to 2.3 \mum CO bands observed by
UKIRT in July, 1997 (see Fig. \ref{l6}) which is consistent with
VLTP (the star on very late thermal pulse) interpretation of V 4334
Sgr (Herwig 2001).

%%%% Figure 6.

\section{Conclusions}

Due to the high sensitivity of computed model atmospheres and
synthetic spectra of evolved stars on the adopted input data, the
analysis of their spectra should be done in the framework of the
self-consisted approach. Model atmospheres should correspond with
the obtained abundances and other physical parameters. 
Even in the case of the ``normal'' red giants the temperature structure
should be computed with the correct abundances.
Otherwise, the abundance determination results might
be affected by significant errors ($>$ 0.2-0.3 dex, see Pavlenko
\& Yakovina 1994 for more details).
Generally
speaking, existed grids of the model atmospheres can be used as
zero-approach tool to be refined in the process of analysis.

Despite a substantial efforts to develop more sophisticated model
atmosphere, a few problems provide the real challenge for the
modern theory of stellar astrophysics:

-- atmospheres of the most evolved stars do not exist in
hydrostatic equilibrium. In some cases effects of sphericity can
be essential.

-- Convection in the extended photospheres can be properly described only
in the 3D approach.

-- The further computation of line lists and other
physical parameters (dissociation energies, partition functions,
etc) of polyatomic molecules is absolutely essential for cool 
stellar atmosphere modelling to progress.
In particular little, no or only poor data exists for species 
such as C$_3$, CH$_4$ and C$_2$H$_2$. These species 
provide substantial opacity in the coolest atmospheres. 
Even the existing line lists of diatomic molecules, 
such as CN, C$_2$, CH, are not accurate enough to be used for 
the high resolution analysis of observed spectra.

Naturally, these problems can be solved one at a time. However, it is only possible
when paying special 
attention to theoretical support of current observations.

\begin{acknowledgements}
I thank SOC \& LOC of the VIII Torino Meeting for the invitation
and support of my participation. I am grateful to my colleagues
Andrew Longmore (Eddinbough Observatory),
Bogdan Kaminsky (MAO),
Carlos Abia (University of Granada), 
Gregory Harris(UCL), 
Hilmar Duerbeck (Open University),  
Hugh Jones(UH), 
Larisa Yakovina (MAO),
Nye Evans (Keele University), 
Thomas Geballe (UKIRT) 
for the excellent collaboration.

 Thanks to Greg Harris (UCL) for his help with our expression of the  
 English language.

\end{acknowledgements}

\bibliographystyle{aa}

\begin{thebibliography}{}

%%\bibitem[{Boesgaard et al. (1998)}]{boe98}
%%Boesgaard, A.\ M.,  et al. 1998, ApJ, 493, 206


%%\bibitem[Cuoco et al.(2003)]{cuoco} Cuoco, A. et al.\ 2003, ArXiv
%%Astrophysics e-prints, 7213

\bibitem[{Anders \& Grevesse(1989)}]{anders89} Anders, E., Grevesse, N.\  1989,
         GeGoAA, 53, 197


\bibitem[Asplund et al. (1997)]{asp97} Asplund, M.,
Gustafsson, B., Lambert, D.L., Rao, N.K.\
1997, A\&A 321, L17

\bibitem[Chandra et al. (1995)]{Chandra}
Chandra S., Kegel W. H., Le Roy R. J.,
Hertenstein T.\ 1995, AAAS, 114, 175

\bibitem[Dominguez et al. (2004)]{Dom}
Dominguez, I, Abia, C., Straniero, O., Cristallo S. and Ya.V. Pavlenko.\
           2004, A\&A, 422, 1045.


\bibitem[Doyle (1968)]{Doyle1968}
 Doyle, R.O.\ 1968, ApJ, 153, 987.

\bibitem[Duerbeck \& Benetti (1996)]{due96}Duerbeck, H.W., 
Benetti, S.\ 1996, ApJ, 468, L111

\bibitem[Jones et al.(2002)]{jones} Jones, H. R. A., Pavlenko, Ya., Viti, S.,
         Tennyson, J.\ 2002, MNRAS, 330, 675


\bibitem[Hauschildt et al.(1999)]{hauschildt99} Hauschildt, P. H., Allard, F.,
         Baron, E.\ 1999, ApJ, 512, 377

\bibitem[Herwig (2001)]{herwig} Herwig, F.\ 2001, ApJ, 554, L71

\bibitem[Goorvitch (1994)]{goorvitch} Goorvitch, D.\ 1994, ApJS, 95, 535


\bibitem[Harris et al. (2002)]{linepaper} Harris G. J., Polyansky O. L.,
Tennyson J.\ 2002, ApJ, 578, 657

\bibitem[Harris et~al. (2003)]{paper1} Harris G. J.,
Pavlenko, Ya. V., Jones, H. R. A., Tennyson J.\ 2003, MNRAS, 344, 1107

\bibitem[Harris et~al. (2006)]{paper2} Harris G. J., Tennyson J.,
Kaminsky B. M., Pavlenko, Ya. V., Jones, H. R. A.\ 2006, MNRAS, 367, 400.


\bibitem[Nakano et al. (1996)]{nak96}Nakano, S., Sakurai, Y., et al.\ 1996, IAU Circ. 6322

\bibitem[Kamath \& Ashok (1999)]{kam99}Kamath, U.S., Ashok, N.M.\ 1999, MNRAS 302, 512


\bibitem[Kerber et al. (2000)]{ker00} Kerber, F., Palsa, R., K\"oppen, J., Bl\"ocker, T.,
  Rosa, M.R.\ 2000, ESO Messenger, No. 101, 27

\bibitem[Kimeswenger et al. (1997)]{kim97} Kimeswenger, S., Gratl, H., Kerber, F., Fouqu\'e, P.,
  Kohle, S. Steele, S.\ 1997, IAU Circ. 6608


\bibitem[Kipper \& Klochkova (1997)]{kip97} Kipper, T.,
Klochkova, V.\ 1997, A\&A 324, L65

\bibitem[Kupka et al.(1999)]{kupka99} Kupka F., Piskunov, N., Ryabchikova, T.A.,
         Stempels, H. C., Weiss, W. W.\ 1999, A\&AS, 138, 119

\bibitem[Kurucz (1993)]{kurucz93} Kurucz.\ 1993, CD ROM 9, 18.
         Harvard-Smisthonian Observatory.

\bibitem[Kurucz (1999)]{kurucz99}Kurucz, R. L.\ 1999, http://kurucz.harvard.edu


\bibitem[Myerscough \& McDowell (1966)]{Myerscough1966}
Myerscough, V. P., McDowell, M. R. C.\ 1966, MNRAS, 132, 457.

\bibitem[Pavlenko (1999)]{yp1999}
Pavlenko Ya.V.\ 1999. Astron. Rept, 43, 94.

\bibitem[Pavlenko (2000)]{pav00} Pavlenko, Ya.\ 2000, Astron. Rept., 44, 219

\bibitem[Pavlenko \& D\"{u}rbeck(2002)]{phwd}
Pavlenko, Ya. V., Duerbeck, H. W.\ 2001a,
       A\&A, 367, 933.

\bibitem[Pavlenko \& D\"{u}rbeck(2002)]{pav-hwd} Pavlenko, Ya. V., D\"{u}rbeck, 
H. W.\
         2001, A\&A, 367, 993

\bibitem[Pavlenko \& Geballe(2002)]{pav-trg} Pavlenko, Ya. V., Geballe, T. R.\ 2002,
         A\&A, 390, 621

\bibitem[Pavlenko \& Jones(2002)]{pav-jones} Pavlenko, Ya. V., Jones, H. R. A.\
         2002, A\&A, 396, 967



\bibitem[Pavlenko \& Zhukovska (2003)]{PZ2003}
Pavlenko, Ya,V., Zhukovska, S.\ 2003,
          KFNT, 19, 28.


\bibitem[Pavlenko et al.(2003)]{pavlenko03b}
         Pavlenko, Ya. V., Jones, H. R. A., Longmore, A. J.\ 2003b,
         MNRAS, 345, 311.

\bibitem[Pavlenko et al.(2003)]{P93} Pavlenko, Ya,V.\ 2003,  
            Astron. Rept., 47, 59.


\bibitem[Pavlenko et al. (2004)]{pav04}
Pavlenko, Ya.V., Geballe, T.R, Evans, A., Smalley, B., Eyres, S.P.S.,
          Tyne, V.H., Yakovina, L.A.\ 2004, A\&A, 417, L39.


\bibitem[Patridge \& Schwenke (1998)]{_PS1998_} Partrige, H., 
                                        Schwenke, D.J.\ 1997,
                          Chem. Phys. 106, 4618.

\bibitem[Plez (1889)]{Plez1989_} Plez, B.\ 1998, A\&A, 337, 495.


\bibitem[Seaton et al.(1992)]{seaton} Seaton M.J., Zeippen C.J., Tully J.A.,
         et al.\ 1992, Rev. Mexicana Astron. Astrophys., 23, 107.

\bibitem[Sneden et~al.(1976)]{Sneden1976}
Sneden, C., Johnson, H. R., Krupp, B. M.\ 1976, ApJ, 204, 218.


\bibitem[Unsold (1955)]{Unsold1955} Unsold, A.\ 1955. Physik der 
Sternatmospharen, springer Verlag, 2-nd ed. 

\bibitem[Yakovina et al. (2003)]{yak2003}
Yakovina, L. A., Pavlenko, Ya. V., Abia, C.\ 2003,
          Ap\&SS, 288, 279.



\end{thebibliography}

\end{document}